# The Kondo Temperature of SU(4) Symmetric Quantum Dots

Michele Filippone[1,2], Cătălin Pașcu Moca[3,4], Gergely Zaránd[3], and Christophe Mora[1]

[1] *Laboratoire Pierre Aigrain, École Normale Supérieure,
Université Paris 7 Diderot, CNRS; 24 rue Lhomond, 75005 Paris, France*

[2] *Dahlem Center for Complex Quantum Systems and Institut für Theoretische Physik,
Freie Universität Berlin, Arnimallee 14, 14195 Berlin, Germany*

[3] *BME-MTA Exotic Quantum Phase Group, Institute of Physics,
Budapest University of Technology and Economics, H-1521 Budapest, Hungary and*

[4] *Department of Physics, University of Oradea, 410087, Oradea, Romania*

A path integral approach is used to derive a closed analytical expression for the Kondo temperature of the SU(4) symmetrical Anderson model. In contrast to the SU(2) case, the prefactor of the Kondo temperature is found to display a peculiar orbital energy (gate voltage) dependence, reflecting the presence of various SU(4) mixed valence fixed points. Our analytical expressions are tested against and confirmed by numerical renormalization group computations.

*Introduction* - Much of our perception of strongly interacting fermion systems such as heavy fermions [1], molecules and quantum dots attached to electrodes [2], correlated states of interacting cold atoms [3], or phenomena such as the Mott transition [4] relies on the detailed study of magnetic impurities and the Kondo effect. The latter consists of the dynamical screening of a local spin entity, interacting antiferromagnetically with surrounding itinerant electrons [5], and provides the simplest example of asymptotic freedom. The fundamental energy scale, at which the Kondo screening develops is called the Kondo temperature, $T_\text{K}$. Our understanding of this curious effect has been strengthened by several recent experiments on quantum dot (QD) systems and nanotubes, which not only allow an accurate control of the model parameters, but also enable one to study out-of-equilibrium phenomena [2, 5, 6] and to realize exotic correlated states such as the two-channel Kondo state [7] or the SU(4) Kondo effect [6, 8, 9] in a controlled manner.

One of the most fascinating features of the Kondo effect is that of *universality*: apart from a few dimensionless numbers characterizing the asymmetry of the leads and electron-hole symmetry, at low temperatures or voltages, every physical quantity depends on the microscopic model parameters solely through $T_\text{K}$. While determining $T_\text{K}$ is therefore clearly of crucial importance, estimating it precisely is an utmost delicate problem. Its first estimate is due to Kondo [10], who constructed and studied a model of a local spin **S** (the spin on the QD or the molecule) coupled to the lead (conduction) electrons' spin density through an exchange coupling, $H_\text{int} = J_0\, \mathbf{S} \cdot \mathbf{s}$. Kondo's exchange model contains, however, logarithmic singularities, which must be regularized by a bandwidth cut-off, $\mathcal{D}_0$. The Kondo temperature is found to depend explicitly on this unknown parameter, and is expressed as $T_{\text{K,SU(2)}} \approx \mathcal{D}_0 \sqrt{\nu_0 J_0} \exp\left(-1/\nu_0 J_0\right)$ for SU(2) spins [5], with $\nu_0$ denoting the electrons' density of states. Unfortunately, $\mathcal{D}_0$ and $J_0$ being both somewhat arbitrary, the predictive power of this expression is limited.

A better estimate can be obtained through the analysis of Anderson's more elaborate impurity model [5], as first done by Haldane [11, 12]. In Anderson's model, some local fermionic degrees of freedom $d_\tau$ of energy $\varepsilon_d$ interact with each other by a local interaction of strength $U$ and hybridize with the leads (conduction electrons) as described by the Hamiltonian,

$$\begin{aligned}H_\text{An} =& \varepsilon_d \sum_\tau d_\tau^\dagger d_\tau + U \sum_{\tau<\tau'} n_\tau n_{\tau'} \\ &+ t \sum_{k\tau}\left(c_{k\tau}^\dagger d_\tau + \text{h.c.}\right) + \sum_{k,\tau} \varepsilon_k c_{k\tau}^\dagger c_{k\tau}\,.\end{aligned} \quad (1)$$

Here the operators $c_{k\tau}^\dagger$ create conduction electrons of energy $\varepsilon_k$ and 'spin' $\tau$, and the strength of hybridization is characterized by the decay rate $\Delta = \pi \nu_0 t^2$. In contrast to the Kondo model, the Anderson model is *not* ultraviolet divergent, and has a well-defined Kondo scale in terms of its three parameters, $\varepsilon_d$, $U$, and $\Delta$, which Haldane determined accurately through a renormalization group analysis for the spin 1/2 case, $\tau = \uparrow, \downarrow$. A closer analysis of Haldane's work reveals, however, that, to provide a precise estimate of the Kondo temperature, one also needs to account for charge (valence) fluctuations.

In this work, we study the SU(4) symmetrical Anderson model (with $\tau = 1, .., 4$), and determine its Kondo temperature accurately. The study of the SU(4) case is motivated by its particular experimental relevance: signs of an emergent SU(4) symmetrical Kondo state were observed in carbon nanotube [13] and vertical quantum dots [9], single atom transistors [14], double QD devices [6, 15], and other exotic SU(4) Kondo regimes are also expected in fourfold degenerate QD systems where both spin and orbital degrees of freedom are conserved in tunneling [15]. Here we develop a path integral formalism introduced in Ref. [16] (see also Ref. [17] for a similar approach), to show that, in contrast to the SU(2) case, the prefactor of the Kondo temperature has a curious dependence on the level position, $\varepsilon_d$. This dependence is a fingerprint of the presence of various SU(4) mixed valence fixed points, governing the charge transitions of the QD (d-level). Our analytical predictions for $T_K$ are found to agree quantitatively with Numerical Renormalization Group (NRG) computations, which also confirm the presence and importance of SU(4) mixed valence fixed points [18].

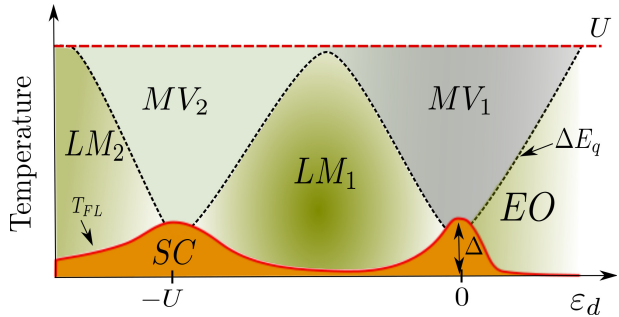

FIG. 1. Sketch of the 'phase diagram' of the SU(4) Kondo model. In the mixed valence region $MV_1$ ($MV_2$) the fourfold degenerate $q = 1$ state competes with the $q = 0$ empty state (the sixfold degenerate $q = 2$ state). In the local moment region $LM_1$ the $q = 1$ SU(4) spin produces Kondo effect and is screened below the Kondo (Fermi liquid) scale, $T_K$ ($T_{FL}$), while in region $LM_2$ a sixfold degenerate SU(4) spin is screened.

The corresponding mixed valence regions and the 'phase diagram' of the SU(4) Kondo model are sketched in Fig. 1 [19]. In the mixed valence regions $MV_1$ and $MV_2$ the fourfold degenerate charge $q = 1$ state of the QD (d-level) competes with the $q = 0$ empty and the sixfold degenerate $q = 2$ states, respectively. The SU(4) spins are formed at temperatures below the energy of the QD's charging excitations $\Delta E_q$ (in local moment regions $LM_1$ and $LM_2$), and get screened below the Kondo (Fermi liquid) scale $T_K$ ($T_{FL}$). The scale $\Delta E_q$ vanishes, however, as one approaches the valence transition points, $\varepsilon_d \approx 0$ and $\varepsilon_d \approx -U$, where the mixed valence fixed points determine the physics down to the Fermi liquid scale, $T_{FL} \sim \Delta$.

To determine $T_K$, we shall focus on the Kondo regime of the QD, with $q = 1, 2$ and $3$ electrons trapped on the fourfold degenerate level, and establish a mapping between the SU(4) Anderson model and the SU(4) Kondo model [20], described by the exchange interaction, $H_{\text{exch}}^{\text{SU(4)}} = \frac{J}{2} \sum_{kk',\tau\tau'} \left( d_{\tau'}^\dagger d_\tau - 1/4 \right) c_{k\tau}^\dagger c_{k'\tau'}$. We determine first $J$ and the cut-off $\mathcal{D}$ by computing the logarithmic corrections to the amplitude of the exchange processes at energy $\omega$

$$\mathcal{J}(\omega) = J - \frac{N\nu_0}{2} J^2 \ln \frac{\omega}{\mathcal{D}} + \dots \quad (2)$$

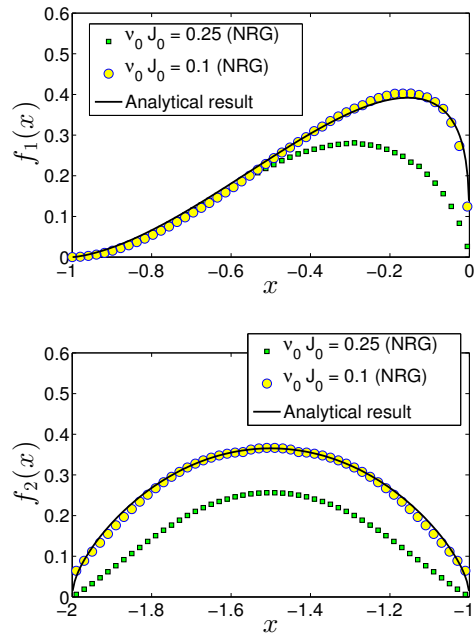

FIG. 2. Comparison of the analytical prefactors $f_1(x)$ and $f_2(x)$ in Eq. (4) (solid lines) with $\mathcal{D}_{\text{eff}}/U$, extracted from NRG calculations (symbols). The numerical calculations were carried out for $\nu_0 J_0 = \{0.1, 0.25\}$. The sector $q = 3$ is symmetric to that with $q = 1$: $f_3(x - 3) = f_1(-x)$.

in both models. Using then the well-known two-loop result for the Kondo temperature, $T_{K,\text{SU(N)}} = \mathcal{D} \sqrt[N]{\frac{N\nu_0 J}{2}} e^{-\frac{2}{N\nu_0 J}}$, we obtain the Kondo temperature of the SU(4) Anderson model as

$$T_{K,\text{SU(4)}} = \mathcal{D}_{\text{eff}} \sqrt[4]{2\nu_0 J_0}\, e^{-\frac{1}{2\nu_0 J_0}}, \quad (3)$$

$$\mathcal{D}_{\text{eff}} = U f_q(\varepsilon_d/U), \quad (4)$$

with $J_0 = -2t^2 U/[E_q(E_q + U)]$ the usual leading order expression for the Kondo coupling, expressed in terms of the shifted energies $E_q = \varepsilon_d + (q-1)U$ [20]. Eq. (3) differs from the SU(2) formula derived by Haldane in that the prefactor in Eq. (3) depends explicitly on the orbital energy (gate voltage) $\varepsilon_d$ through the function $f_q$, specified later [21]. We emphasize that Eq. (3) incorporates the effects of valence fluctuations, which have a determining impact on the functions $f_q$.

To verify the analytical predictions, we also performed NRG computations using the Budapest Flexible DM-NRG code [22], and compared the numerically obtained Kondo temperature to the analytical result, Eq. (4). As shown in Fig. 2, an excellent agreement is found for small values of $\nu_0 J_0$ [23]. Clearly, our results contradict the naive expectation that $\mathcal{D}$ should roughly correspond to the minimum charging energy of the quantum dot, in which case the edges of the functions $f_q$ would exhibit
2

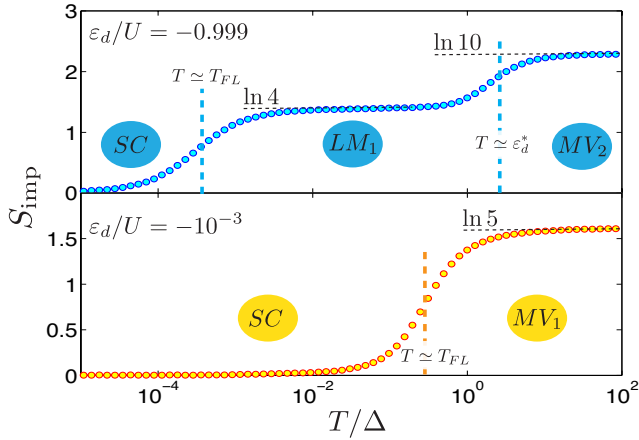

FIG. 3. Temperature dependence of impurity contribution to the entropy.

a linear behavior. Rather, as discussed below, for small $\Delta$'s the value of $\mathcal{D}_{\text{eff}}$ is strongly renormalized by valence fluctuations. While for $f_2$, the discrepancy with linearity close to $x = -2$ and $-1$ is not very strong, it is much more pronounced for $f_1$ where, in addition, a strong asymmetry is observed. These differences are the results of the competition of different charge configurations in the vicinity of the mixed valences regions $MV_1$ and $MV_2$ (see Fig. 1), as we discuss next.

*Valence fluctuations* - The importance of charge fluctuations is evident from the numerical study of the SU(4) Anderson model. Fig. 3 displays the temperature dependence of the impurity entropy for two values of $\epsilon_d$ close to the $q = 0 \to 1$ and $q = 1 \to 2$ charge transitions, as determined by our DM-NRG computations. The impurity entropy $S_{\text{imp}}(T)$ displays steps at each temperature corresponding to a 'phase boundary' in Fig. 1. For $\epsilon_d \approx -0.001 U$, e.g., the impurity entropy takes a value of $S_{\text{imp}} = k_B \ln 5$ at high temperatures, corresponding to the approximate (1+4)-fold degeneracy of the $q = 0$ and $q = 1$ states (region $MV_1$), and by decreasing the temperature further below $T_{FL}$, one enters directly the Fermi liquid regime with $S_{\text{imp}} = 0$. For $\varepsilon_d \approx -0.999 U$, on the other hand, the entropy takes a value of $k_B \ln 10$ at hight temperatures, reflecting the (6+4)-fold degeneracy of the almost degenerate $q = 2$ and $q = 1$ charging states (region $MV_2$), and a reduction series of $\ln 10 \to \ln 4 \to 0$ is observed, corresponding to the $MV_2 \to LM_1$ and $LM_1 \to$ 'Fermi liquid' transitions.

From the previous analysis it is clear that, in the vicinity of the charging transitions, the mixed valence fixed points govern the physics over a large energy window, and have a substantial impact on quantum fluctuations - determining the precise Kondo scale. In fact, the basic structure of the functions $f_q$ can be understood by means of a relatively simple renormalization group procedure constructed by Haldane [11], accounting also for charge fluctuations. Before proceeding with the full derivation of the functions $f_q$, let us present this simple and instructive derivation. We first consider the case $\Delta \ll -\varepsilon_d \ll U$ where the $q = 0$ and $q = 1$ states compete in region $MV_1$. In a standard renormalization group (RG) picture, $\varepsilon_d$ is first renormalized by charge fluctuations in the region $MV_1$, and replaced by

$$\varepsilon_d^* = \varepsilon_d + \frac{3\Delta}{\pi} \ln \frac{U}{\alpha|\varepsilon_d|} \quad (5)$$

in the region $LM_1$, with $\alpha$ a constant of order one. The factor 3 in Eq. (5) accounts for the balance between the empty state and the fourfold degenerate singly occupied state. In region $LM_1$ (below a cutoff energy $\mathcal{D} \approx \alpha\varepsilon_d$), the QD can only be singly occupied, and a standard SU(4) Kondo effect takes place. The exchange constant $J$ can be estimated here by the Schrieffer-Wolff approach as $J = -2\Delta/(\pi\varepsilon_d^*)$. Combining $\mathcal{D}$ and $J$ within the two-loop expression of the Kondo temperature, we recover Eqs. (3) with $\mathcal{D}_{\text{eff}} = \sqrt[4]{\alpha|\varepsilon_d|}U^{3/4}$, yielding $f_1(x)|_{x\to 0^-} \propto \sqrt[4]{|x|}$ in agreement with the analytical expression of $f_1(x)$ (with $\alpha^{-1} = \gamma_e$ identified as Euler's constant). The reason for the emerging non-integer power is the logarithmic correction to $\varepsilon_d$, Eq. (5), which – after exponentiation in the Kondo temperature's expression – rescales the effective cutoff $\mathcal{D}_{\text{eff}}$ to be of a non-trivial power law form.

The same reasoning can be extended to the proximity of the region $MV_2$. Note that there are six different states with double occupancy. Each of them is coupled to two singly occupied states by the tunneling term, while each singly occupied state is coupled to three states with double occupancy. As a result we find $\mathcal{D}_{\text{eff}}/U = \alpha^{5/4}(\varepsilon_d/U + 1)^{5/4}$ for $\varepsilon_d > -U$ and $\mathcal{D}_{\text{eff}}/U = (\alpha|\varepsilon_d/U + 1|)^{3/4}$ for $\varepsilon_d < -U$ in agreement with the general expressions obtained for $f_1(x)$ and $f_2(x)$ (with again $\alpha = 1/\gamma_e$). The three limiting expressions derived so far by simple scaling arguments are sufficient to qualitatively understand the shape of the effective cutoff $\mathcal{D}_{\text{eff}}$ shown in Fig. 2, in particular the strong asymmetry in $f_1$.

*Path integral approach* - To determine the functions $f_q$ precisely, we must go beyond the previous simple analysis. The Schrieffer-Wolff (SW) transformation, applied to Eq. (1), gives the Kondo coupling $J_0$ of Eq. (3), but only a rough estimate of the cutoff $\mathcal{D}_{\text{eff}}$. Moreover, any subleading correction to $J_0$ renormalizes $\mathcal{D}_{\text{eff}}$. A precise determination of $\mathcal{D}_{\text{eff}}$ thus requires a second order calculation in $\Delta$, or fourth order in the tunneling $t$, beyond SW. We follow the path integral approach devised in Ref. [16], where a systematic expansion in $\Delta$ is performed, and extend it to the SU(4) symmetry. Bosonic (fermionic) slave fields, associated with each even (odd) charged state of the QD, are introduced. They give a quadratic Anderson Hamiltonian at $t = \Delta = 0$ around which a diagrammatic perturbation theory in $\Delta$ is used. Below, we focus on the case $q = 1$, but a similar discussion applies equally for the cases $q = 2$ and $q = 3$.

In the path integral formalism, the high energy slave-fields, associated with QD states with $q \neq 1$, are integrated out, and the Anderson model is mapped onto a SU(4) Kondo-like action

$$S_{\rm K} = \sum_{\tau\tau'=1}^{4} {\rm Tr}\left[\frac{\tilde{J}}{2}\left(f_{\tau'}^\dagger f_\tau - \frac{1}{4}\right) c_{k\tau}^\dagger c_{k'\tau'}\right]. \quad (6)$$

The trace in Eq. (6) refers to summations over all wavevectors and Matsubara frequencies. In contrast to the genuine Kondo model, $\tilde{J}$ in Eq. (6) is a frequency dependent coupling [16]. This dependence is a remainder of the integrated charge fluctuations in the Anderson model in addition to the Kondo-like spin fluctuations. To leading order in $\Delta$ and neglecting its frequency dependence, $\tilde{J}$ reduces to the SW coupling $J_0$. For energies $\omega$ below the charging energy $U$, dot excitations out of the $q = 1$ subspace are frozen and the Anderson model maps (for $-\varepsilon_d \gg \Delta$ and $\varepsilon_d + U \gg \Delta$) onto the Kondo model. The Kondo temperature $T_{\rm K}$ becomes the only relevant energy scale, which controls the Kondo crossover. As a result, the expansion of the spin-exchange process $\mathcal{J}(\omega)$ in the Anderson model recovers exactly the same expansion as Eq. (2), derived within the Kondo model (see Eq. (7a)). We can thus identify $J$ and $\mathcal{D}$ from this weak coupling expansion, and determine $T_{\rm K}$.

Equipped with the action Eq. (6), we calculate the spin-exchange process $\mathcal{J}(\omega)$, given by the series of diagrams represented in Fig. 4. To be consistent with the second order expansion in $\Delta$, self-energy vertex corrections must also be considered. They lead to a renormalization of the orbital energy, analogous to Eq. (5), $\varepsilon_d \to \tilde{\varepsilon}_d = \varepsilon_d + \nu_0 t^2 \left[\ln(-\varepsilon_d)/\Xi + 3\ln(\varepsilon_d + U)/\Xi\right]$, with $\Xi$ denoting an UV regularization cutoff (to be distinguished from $\mathcal{D}$), eventually sent to infinity. The self-energy corrections also give rise to a *quasi-particle weight* $\mathcal{Z} = 1 + \nu_0 t^2 \left[1/\varepsilon_d - 3/(\varepsilon_d + U)\right]$ for the slave-fermions $f_\tau$, which renormalizes the spin-exchange process. The complete calculation leads to the renormalized exchange amplitude,

$$\mathcal{J}(\omega) = J_0 + 2\nu_0 J_0^2 g_1\left(\frac{\varepsilon_d}{U}\right) - 2\nu_0 J_0^2 \ln\frac{\omega}{\mathcal{D}}, \quad (7{\rm a})$$

$$g_1(x) = \frac{1}{4}\frac{3x-2}{x+2} - \frac{x^2}{2}\frac{\left(x^2 + 3x + 3\right)}{(x+2)^2} \ln\frac{2x+3}{x+1}, \quad (7{\rm b})$$

with $\mathcal{D} = \sqrt[4]{-\varepsilon_d(\varepsilon_d + U)^3}$. Comparison with Eq. (2) for $N = 4$ yields then the exchange coupling $J = J_0 + 2\nu_0 J_0^2 g_1(\varepsilon_d/U)$. Finally, computing the Kondo temperature $T_{\rm K,SU(4)}$, we obtain Eq. (3) with

$$f_1(x) \equiv \sqrt[4]{-x(x+1)^3} \, \exp\left[g_1(x)\right]. \quad (8)$$

The function $f_2(x)$ can be derived by a similar procedure. Details of its rather lengthy derivation as well as the analytical form of $f_2$ are given in the Supplemental Material [24].

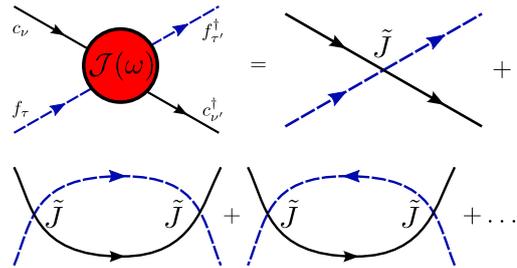

FIG. 4. Diagrammatic expansion from Eq. (6) for the spin-exchange process $\mathcal{J}(\omega)$ to second order in $\Delta$. Scattering lead electrons are considered at the energy $\omega$, while slave fermions at energy $\tilde{\varepsilon}_d$.

*Conclusions* - In this work, we determined the Kondo temperature of the SU(4) Kondo model. In contrast to the SU(2) case, the prefactor of $T_K$ was found to depend strongly on $\epsilon_d$ through universal functions of $\epsilon_d/U$, which we determined analytically and numerically. As demonstrated through an analysis similar to that of Haldane, charge fluctuations play a decisive role in determining the afore-mentioned prefactors, and lead to a power law suppression of the Kondo scale close to the SU(4) mixed valence fixed points. In addition to being of fundamental interest, the anomalous dependence of the Kondo temperature found here should be of importance when characterizing SU(4) Kondo systems.

M.F. acknowledges R. Vasseur and E. Vernier for useful discussions and the Alexander von Humboldt Foundation for partial financial support. G.Z. and C.P.M. acknowledge support from Hungarian Grant Nos. K105149 and CNK80991. C.P.M. was also supported by UEFISCDI under French-Romanian Grant DYMESYS, Contract No. PN-II-ID-JRP-2011-1.

# Supplementary material for the paper "The Kondo Temperature of SU(4) Symmetric Quantum Dots"

Michele Filippone[1,2], Cătălin Paşcu Moca[3,4], Gergely Zaránd[3], and Christophe Mora[1]
[1] Laboratoire Pierre Aigrain, École Normale Supérieure,
Université Paris 7 Diderot, CNRS; 24 rue Lhomond, 75005 Paris, France
[2] Dahlem Center for Complex Quantum Systems and Institut für Theoretische Physik,
Freie Universität Berlin, Arnimallee 14, 14195 Berlin, Germany
[3] BME-MTA Exotic Quantum Phase Group, Institute of Physics,
Budapest University of Technology and Economics, H-1521 Budapest, Hungary and
[4] Department of Physics, University of Oradea, 410087, Oradea, Romania

## S-I. PATH INTEGRAL AND SLAVE BOSON REPRESENTATION OF THE SU(4) ANDERSON MODEL

Here we provide detailed discussion of the techniques developed and the calculations carried out to derive the analytical results in the main text. To describe the Hilbert space of the dot we adopt a slave boson approach. In Fig. S-1, we provide a graphical representation of the quantum dot Hilbert space together with the associated energies and slave-fields. Bosonic operators $b^\dagger$ (fermionic $f^\dagger$) act on a vacuum state $|0\rangle_\lambda$ to create even (odd) charged states: $b_0^\dagger |0\rangle_\lambda$ corresponds to the empty dot $|0\rangle$; $f_\tau^\dagger |0\rangle_\lambda$ to the dot occupied by a single electron with spin $\tau$; $b_{\tau\tau'}^\dagger |0\rangle_\lambda$, with $\tau < \tau'$, corresponds to the 6 possible doubly occupied states $|\tau, \tau'\rangle$; $f_{3,\tau}^\dagger |0\rangle_\lambda$ describes a triply occupied dot in which only the electron in the $|\tau\rangle$ state is missing and $b_4^\dagger |0\rangle_\lambda$ the fully occupied dot. The expression of the original creation operators $d_\tau^\dagger$ in this new basis reads

$$\begin{aligned}
d_1^\dagger &= f_{1,1}^\dagger b_0 + b_{12}^\dagger f_{1,2} + b_{13}^\dagger f_{1,3} + b_{14}^\dagger f_{1,4} + b_4^\dagger f_{3,1} - f_{3,2}^\dagger b_{34} + f_{3,3}^\dagger b_{24} - f_{3,4}^\dagger b_{23}, \\
d_2^\dagger &= f_{1,2}^\dagger b_0 - b_{12}^\dagger f_{1,1} + b_{23}^\dagger f_{1,3} + b_{24}^\dagger f_{3,4} + b_4^\dagger f_{3,2} + f_{3,1}^\dagger b_{34} - f_{3,3}^\dagger b_{14} + f_{3,4}^\dagger b_{13}, \\
d_3^\dagger &= f_{1,3}^\dagger b_0 - b_{13}^\dagger f_{1,1} - b_{23}^\dagger f_{1,2} + b_{34}^\dagger f_{1,4} + b_4^\dagger f_{3,3} - f_{3,1}^\dagger b_{24} + f_{3,2}^\dagger b_{14} - f_{3,4}^\dagger b_{12}, \\
d_4^\dagger &= f_{1,4}^\dagger b_0 - b_{14}^\dagger f_{1,4} - b_{24}^\dagger f_{1,2} - b_{34}^\dagger f_{1,3} + b_4^\dagger f_{3,4} + f_{3,1}^\dagger b_{23} - f_{3,2}^\dagger b_{13} + f_{3,3}^\dagger b_{12}.
\end{aligned} \quad \text{(S-1)}$$

The choice of the signs is fixed to preserve the fermionic anticommutation relations $\{d_\tau^\dagger, d_{\tau'}\} = \delta_{\tau\tau'}$ and $\{d_\tau^\dagger, d_{\tau'}^\dagger\} = 0$ between the original operators. In the new basis, the action associated to the Anderson Hamiltonian Eq. (1) becomes quadratic for $t = \Delta = 0$

$$S_0 = -\int_0^\beta d\tau \left\{ \sum_{k\nu} c_{k\nu}^\dagger(\tau) G_k^{-1} c_{k\nu}(\tau) + b_0^\dagger(\tau) F_0^{-1} b_0(\tau) + \sum_\nu f_{1,\nu}^\dagger(\tau) F_1^{-1} f_{1,\nu}(\tau) + \right. \\ \left. + \sum_{\nu<\nu'} b_{\nu\nu'}^\dagger(\tau) F_2^{-1} b_{\nu\nu'}(\tau) + \sum_\nu f_{3,\nu}^\dagger(\tau) F_3^{-1} f_{3,\nu}(\tau) + b_4^\dagger(\tau) F_4^{-1} b_4(\tau) \right\}, \quad \text{(S-2)}$$

with $G_k = (-\partial_\tau - \varepsilon_k)^{-1}$ the lead electron free propagator and $F_q = (-\partial_\tau - E_q)^{-1}$ the slave-field free propagator, with $E_q = q\varepsilon_d + q(q-1)U/2 + \lambda$. Note that the para-energy $\lambda$ must be introduced in free slave-state propagators to perform Abrikosov's projection onto the physical sector [1], defined by the constraint

$$N_{\text{slave}} = b_0^\dagger b_0 + \sum_\tau f_{1,\tau}^\dagger f_{1,\tau} + \sum_{\tau<\tau'} b_{\tau\tau'}^\dagger b_{\tau\tau'} + \sum_\tau f_{3,\tau}^\dagger f_{3,\tau} + b_4^\dagger b_4 = 1. \quad \text{(S-3)}$$

As $[N_{\text{slave}}, H] = 0$, the Anderson Hamiltonian is diagonal in sectors with integer $N_{\text{slave}}$. $\text{Tr}_n$ is the trace carried out in the subspace defined by $N_{\text{slave}} = n$, $n = 1$ being the physical one. Denoting $\langle \mathcal{O} \rangle_\lambda = \text{Tr}[\mathcal{O} e^{-\beta H - \beta N_{\text{slave}}}]/\text{Tr}[e^{-\beta H - \beta N_{\text{slave}}}]$ the expectation value of the operator $\mathcal{O}$ in the whole Hilbert space spanned by the action Eq. (S-2), the physical average $\langle \mathcal{O} \rangle = \text{Tr}_1[\mathcal{O} e^{-\beta H}]/\text{Tr}_1[e^{-\beta H}]$ is recovered with the following prescription [2]

$$Z = \lim_{\lambda \to \infty} \frac{\partial}{\partial e^{-\beta\lambda}} Z_\lambda, \qquad \langle \mathcal{O} \rangle = \lim_{\lambda \to \infty} \left[ \langle \mathcal{O} \rangle_\lambda + \frac{Z_\lambda}{Z} \frac{\partial}{e^{-\beta\lambda}} \langle \mathcal{O} \rangle_\lambda \right]. \quad \text{(S-4)}$$

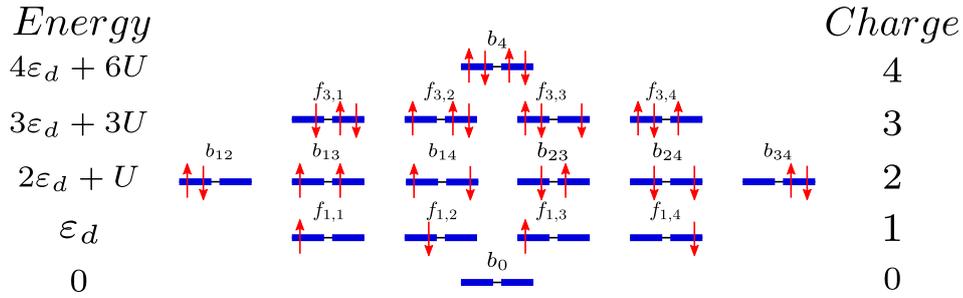

FIG. S-1: Representation of the SU(4) quantum dot Hilbert space with slave-boson and para-fermions. In the case of a SU(4) symmetric carbon nanotube, the label $\tau = 1, \ldots, 4$ merges the fourfold valley and spin degeneracy. This leads to 12 possible states, which are here represented with their energy and total charge occupation. To every configuration a slave-field is associated: bosons for even occupancy and fermions for odd occupancy of the quantum dot.

$Z = \text{Tr}_1\left[e^{-\beta H}\right]$ is the *physical* partition function of the Anderson Hamiltonian. The tunneling term reads

$$S_T = t \int_0^\beta d\tau \sum_{k\nu} \left( c_{k\nu}^\dagger(\tau) d_\nu(\tau) + d_\nu^\dagger(\tau) c_{k\nu}(\tau) \right), \tag{S-5}$$

in which the transformation Eq. (S-1) has to be operated.

The path integral representation permits to map the SU(4) Anderson model onto the SU(4) Kondo action Eq. (6) and carry out the diagrammatic calculations of the exchange processes $\mathcal{J}(\omega)$ in the three charge sectors $q = 1, 2, 3$. The procedure can be summarized in the following way. The slave-fields associated to the high energy charge sectors relevant to fourth order in $t$, see Fig. S-2, are integrated. The effective interaction assumes always a Kondo-like form. In the Matsubara representation of fields, this reads

$$S_{\text{SU}(4)} = \frac{1}{\beta} \sum_{\substack{kk'\nu\nu'\tau\tau' \\ n,m,l}} c_{k\nu}^\dagger(i\omega_m) c_{k'\nu'}(i\omega_l) f_\tau^\dagger(i\nu_n + i\omega_l) f_{\tau'}(i\nu_n + i\omega_m) \left[ \tilde{J}_{n,m,l} \mathbf{T}_{\tau\tau'} \cdot \mathbf{t}_{\nu\nu'} + \mathbb{1}\, \tilde{W}_{n,m,l} \right]. \tag{S-6}$$

$\mathbf{T}$ and $\mathbf{t}$ are the vectors collecting the 15 matrices forming the fundamental representation of the SU(4) group [3], providing a compact notation of the SU(4) exchange interaction

$$H_{\text{exch}}^{\text{SU}(4)} = \frac{J}{2} \sum_{kk',\tau\tau'} \left( f_{\tau'}^\dagger f_\tau - \frac{q}{4} \right) c_{k\tau}^\dagger c_{k'\tau'} = J \sum_{\nu\nu'\tau\tau'} \mathbf{T}_{\tau\tau'} \cdot \mathbf{t}_{\nu\nu'} c_\nu^\dagger c_{\nu'} f_\tau^\dagger f_{\tau'}, \tag{S-7}$$

for a dot occupancy $\sum_\tau f_\tau^\dagger f_\tau = q$. Notice that a potential scattering term appears in Eq. (S-6). This is neglected in the main text as it does not affect the nature of the Fermi liquid fixed points [4]. The antiferromagnetic $\tilde{J}$ and the potential scattering $\tilde{W}$ are frequency-dependent couplings [4], in contrast with the genuine Kondo model. This is a reminder of the charge fluctuations in the Anderson model in addition to the Kondo-like spin fluctuations. We focus to the case of single dot occupancy $q = 1$. The following reasoning can be extended to $q = 2$ and 3, as discussed in Sections S-III and S-IV. For energies $\omega$ below the charging energy $U$, dot excitations towards the doubly occupied and empty states are frozen and the Anderson model maps rigorously (for $-\varepsilon_d \gg \Delta$ and $\varepsilon_d + U \gg \Delta$) onto the Kondo model. The Kondo temperature $T_K$ becomes the only relevant energy scale, controlling the evolution from weak coupling (high energy) to strong coupling (low energy). As a result, we expect the expansion of the spin-exchange coupling $\mathcal{J}(\omega)$ in the Anderson model to recover exactly the same expansion derived in the Kondo model Eq. (2). The identification at weak coupling ($T_K \ll \omega \ll U$) is then sufficient to determine $J$, $\mathcal{D}$ in Eq. (2), and therefore $T_K$.

The above statements can be framed more rigorously in a renormalization group approach. The integration over the fields with energies $\varepsilon_k$ larger than the energy scale $\omega$ leads to an effective action for the system [5, 6]. The fields describing lead electrons and slave-fermions in Eq. (S-6) are renormalized by interactions upon varying $\omega$. Their propagators are replaced by the full propagators $\mathcal{G}_k^\omega$ and $\mathcal{F}_1^\omega$ with Dyson equations, e.g. $\mathcal{F}_1^\omega(i\omega_n) = \left[ F_1(i\omega_n)^{-1} - \Sigma_1^\omega(i\omega_n) \right]^{-1}$, $\Sigma_1^\omega$ being the self-energy. The superscript $\omega$ indicates that all internal lines in diagrams contributing to $\Sigma^\omega$ are integrated for momenta $k$ such that $\omega \leq \varepsilon_k \leq \Xi$. $\Xi$ is an intermediate cutoff introduced to regularize UV divergences, and safely sent to infinity at the end of calculation. The interaction term in Eq. (6) is replaced in the

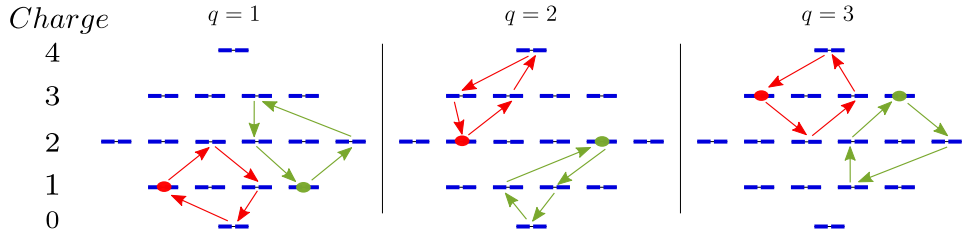

FIG. S-2: Virtual paths in the SU(4) Anderson model to fourth order in the tunnel coupling $t$. We consider different Coulomb blockade regimes with $q$ charges in the quantum dot. The circles signal the departure charge state to which one has to come back in perturbation theory. We emphasize graphically the symmetry between the sectors $q=1$ and 3. Notice that, for $q=1$ and 3, the charge sectors 4 and 0 can be respectively neglected to fourth order in $t$.

course of renormalization by the four-point vertex function $\mathcal{V}^\omega$ [7]

$$\mathcal{V}^\omega_{\nu\nu',\tau\tau'}(k,i\omega_l,i\omega_m;k'i\omega_n,i\omega_o) = -\beta \frac{\left\langle c_{k\nu}(i\omega_l)g_\tau(i\omega_m)g^\dagger_{\tau'}(i\omega_o)c^\dagger_{k'\nu'}(i\omega_n)\right\rangle\big|_c}{G_k(i\omega_l)G_{k'}(i\omega_n)F_q(i\omega_m)F_q(i\omega_o)}. \tag{S-8}$$

Renormalization also generates higher order vertices, which are irrelevant and discarded from the effective action. Before writing the resulting effective action, let us proceed with a simple rescaling of fields. The slave-fermion propagator $\mathcal{F}^\omega_1$ is expanded close to its *renormalized pole* $\tilde{\varepsilon}_d = \varepsilon_d + \lambda + \Sigma^\omega_1(\varepsilon_d)$ so we can write $\mathcal{F}^\omega_1(i\omega_n) = \mathcal{Z}/(i\omega_n - \tilde{\varepsilon}_d)$. $\mathcal{Z} = \left(1 - \partial_{i\omega_n}\Sigma(i\omega_n)|_{i\omega\to\tilde{\varepsilon}_d}\right)$ is the *quasi-particle weight* of the slave-fermions, corresponding to the renormalization of their wavefunction. Rescaling the fields $f^{(\dagger)}_\tau \to \tilde{f}^{(\dagger)}_\tau = f^{(\dagger)}_\tau/\sqrt{\mathcal{Z}}$ and expanding the vertex function $\mathcal{V}^\omega$ close to the propagator poles $E_F = 0$ (the Fermi energy) and $\tilde{\varepsilon}_d$, the effective action of the system is derived

$$S' = -\sum_{k,i\omega_n,\tau}^{<\omega} c^\dagger_{k\tau}G^{-1}_k c_{k\tau} - \sum_{i\omega_n,\tau}^{<\omega} \tilde{f}^\dagger_\tau \tilde{F}^{-1}_1 \tilde{f}_\tau + \sum_{kk'i\omega\tau\tau'\nu\nu'}^{<\omega} \mathcal{V}^R_{\tau\tau',\nu\nu'}c^\dagger_{k\nu}c_{k\nu'}\tilde{f}^\dagger_\tau\tilde{f}_{\tau'} + \ldots \tag{S-9}$$

$\tilde{F}_1 = (i\omega_n - \tilde{\varepsilon}_d)^{-1}$ has the same form as the free propagator $F_1$. $\mathcal{V}^R = \mathcal{Z}\mathcal{V}^\omega$ is the *renormalized vertex* of the theory [8]. We neglect any frequency dependent correction to the action Eq. (S-9) as they are irrelevant in the renormalization flow. The renormalized vertex reads [9–12]

$$\mathcal{V}^R = \mathcal{J}(\omega)\mathbf{T}\cdot\mathbf{t} + \mathcal{W}\mathbb{1}. \tag{S-10}$$

This quantity involves the spin-exchange process $\mathcal{J}(\omega)$, which recovers the universal expression Eq. (2) in the main text, valid for a pure SU(4) Kondo model. The couple $(J,\mathcal{D})$ is then set and the Kondo temperature determined through $T_{K,\mathrm{SU(N)}} = \mathcal{D}\sqrt[N]{N\nu_0 J/2}\; e^{-\frac{2}{N\nu_0 J}}$, leading to our main result Eqs. (3) and (4). In the following, we give a comprehensive discussion of the calculation for the spin-exchange process for dot occupancy $q=1$. The main differences with the other charge sectors $q=2$ and 3 are also pointed out.

## S-II. CALCULATION OF THE SU(4) SPIN EXCHANGE PROCESSES FOR $q=1$

### 1. Integration of the high energy slave-fields

The quantum dot hosts a single charge if $-U < \varepsilon_d < 0$. In Fig. S-2 we show that virtual processes to fourth order in the tunneling $t$ do not involve the charge sector occupied by four charges. The associated slave-field $b_4$ can then be neglected. The action of the system reduces to $S = S_0 + S_{01} + S_{12} + S_{23}$, where $S_0$ is the quadratic action Eq. (S-2) without the $b_4$ field and $S_{nm}$ hybridizes the slave-fields associated to $n$ and $m$ charges on the quantum dot. For instance, $S_{01}$ reads

$$S_{01} = t\,\mathrm{Tr}\left[b^\dagger_0 \sum_{k\tau} c^\dagger_{k\tau}f_{1,\tau} + \mathrm{h.c.}\right]. \tag{S-11}$$

The trace refers equally to the summation over the imaginary-time or Matsubara frequency dependence of the fields, *e.g.*

$$\text{Tr}\left[c_k^\dagger G_k^{-1} c_k\right] = \int_0^\beta d\tau c_k^\dagger(\tau)(-\partial_\tau - \varepsilon_k)c_k(\tau) = \sum_n c_k^\dagger(i\omega_n)(i\omega_n - \varepsilon_k)c_k(i\omega_n). \quad \text{(S-12)}$$

The integration over the $b_0$ slave-field leads to an effective interaction of the form Eq. (S-6)

$$S'_{01} = -\frac{t^2}{\beta} \sum_{\substack{kk'\nu\nu'\tau\tau' \\ n,m,l}} F_0(i\nu_n) c_{k\nu}^\dagger(i\omega_m) c_{k'\nu'}(i\omega_l) f_{1,\tau}^\dagger(i\nu_n + i\omega_l) f_{1,\tau'}(i\nu_n + i\omega_m) \left[2\mathbf{T}_{\tau\tau'} \cdot \mathbf{t}_{\nu\nu'} - \frac{1}{4}\mathbb{1}_{\tau\tau',\nu\nu'}\right], \quad \text{(S-13)}$$

in which we make use of Eq. (S-7) with $q = 1$. The $f_{3,\tau}$ slave-fields are then integrated. The action $S_{23}$ has the algebraic form

$$S_{23} = t\,\text{Tr}\left[\mathbf{F}^\dagger \cdot \mathbf{C} \cdot \mathbf{b}\right] + \text{h.c.} \quad \text{(S-14)}$$

$$\mathbf{F} = \begin{pmatrix} f_{1,1} \\ f_{1,2} \\ f_{1,3} \\ f_{1,4} \end{pmatrix}, \qquad \mathbf{b} = \begin{pmatrix} b_{12} \\ b_{13} \\ b_{14} \\ b_{23} \\ b_{24} \\ b_{34} \end{pmatrix}, \qquad \mathbf{C} = \begin{pmatrix} 0 & 0 & 0 & c_4 & -c_3 & c_2 \\ 0 & -c_4 & c_3 & 0 & 0 & -c_1 \\ c_4 & 0 & -c_2 & 0 & c_1 & 0 \\ -c_3 & c_2 & 0 & -c_1 & 0 & 0 \end{pmatrix}. \quad \text{(S-15)}$$

To keep simple notations, we omit the momentum index $k$ for lead electrons. The integration of the para-fermions $f_{3,\tau}$ leads to the effective interaction

$$S'_{23} = t^2 \,\text{Tr}\left[\mathbf{b}^\dagger \cdot \mathbf{A}_2 \cdot \mathbf{b}\right], \quad \text{(S-16)}$$

with $\mathbf{A}_2 = \mathbf{C}^\dagger \cdot F_3 \mathbb{1}_4 \cdot \mathbf{C}$ and $\mathbb{1}_4$ the $4 \times 4$ identity matrix. This term renormalizes the action involving the $b_{\tau\tau'}$ slave-bosons

$$S_{12} + S'_{23} = -\text{Tr}\left[\mathbf{b}^\dagger \cdot \Phi_2^{-1} \cdot \mathbf{b}\right] + t\,\text{Tr}\left[\mathbf{b}^\dagger \cdot \mathbf{w} + \mathbf{w}^\dagger \cdot \mathbf{b}\right], \quad \text{(S-17)}$$

with

$$\Phi_2^{-1} = F_2^{-1} \mathbb{1}_6 - t^2 \mathbf{A}_2 \qquad \text{and} \qquad \mathbf{w} = \begin{pmatrix} f_{1,2}c_1 - f_{1,1}c_2 \\ f_{1,3}c_1 - f_{1,1}c_3 \\ f_{1,4}c_1 - f_{1,1}c_4 \\ f_{1,3}c_2 - f_{1,2}c_3 \\ f_{1,4}c_2 - f_{1,2}c_4 \\ f_{1,4}c_3 - f_{1,3}c_4 \end{pmatrix}. \quad \text{(S-18)}$$

Eq. (S-17) is quadratic in $\mathbf{b}$ and its integration is also Gaussian, leading to

$$S'_{12} = t^2 \,\text{Tr}\left[\mathbf{w}^\dagger \cdot \Phi_2 \cdot \mathbf{w}\right] = t^2 \,\text{Tr}\left[\mathbf{w}^\dagger \cdot F_2 \mathbb{1}_6 \cdot \mathbf{w}\right] + t^4 \,\text{Tr}\left[\mathbf{w}^\dagger \cdot F_2 \cdot \mathbf{A}_2 \cdot F_2 \cdot \mathbf{w}\right]. \quad \text{(S-19)}$$

In the last equality, we have expanded the operator $\Phi_2$ in $t$. The mapping of the first term of Eq. (S-19) onto the Kondo-like interaction Eq. (S-6) possible by considering the identity $\mathbf{w}^\dagger \cdot \mathbf{w} = \sum_{\nu\nu'\tau\tau'} c_\nu^\dagger c_{\nu'} f_{1,\tau}^\dagger f_{1,\tau'} \left[\frac{3}{4}\delta_{\nu\nu'}\delta_{\tau\tau'} - 2\mathbf{T}_{\tau\tau'} \cdot \mathbf{t}_{\nu\nu'}\right]$, a consequence of the relation $\sum_\tau f_{1,\tau}^\dagger f_{1,\tau} = 1$. This is a reformulation of the constraint Eq. (S-3), once all the slave-states participating to it, with the exception of the para-fermions $f_{1,\tau}$, have been integrated out. The effective action is cast into the form

$$S'_1 = S'_0 + S_{\text{SU}(4)} + t^4 \,\text{Tr}\left[\mathbf{w}^\dagger \cdot F_2 \cdot \mathbf{A}_2 \cdot F_2 \cdot \mathbf{w}\right], \quad \text{(S-20)}$$

with $S'_0$ given by the two first terms of Eq. (S-2). $S_{\text{SU}(4)}$ has the structure of Eq. (S-6), with $\tilde{J} = -2t^2\left[F_2 + F_0\right]$ and $\tilde{W} = (t^2/4)\left[3F_2 - F_0\right]$. Their frequency dependence is explicit in the Matsubara representation of the fields. Adopting the same notations as in Eq. (S-6)

$$\tilde{J}^1_{i\nu_n,i\omega_m,i\omega_l} = -2t^2\left[F_0(i\nu_n) + F_2(i\nu_n + i\omega_m + i\omega_l)\right], \qquad \tilde{W}^1_{i\nu_n,i\omega_m,i\omega_l} = -\frac{t^2}{4}\left[F_0(i\nu_n) - 3F_2(i\nu_n + i\omega_m + i\omega_l)\right]. \quad \text{(S-21)}$$

These two expressions are in agreement with the Schrieffer-Wolff transformation for $q=1$ [13, 14]. This can be seen by substituting the frequencies of Eq. (S-21) with the poles of $G_k$ and $F_1$: $i\nu_n \to \varepsilon_d + \lambda$ and $i\omega_{m,l} \to 0$. The last interaction term in Eq. (S-20) reads

$$t^4 \text{Tr}\left[\mathbf{w}^\dagger \cdot F_2 \cdot \mathbf{A}_2 \cdot F_2 \cdot \mathbf{w}\right] = \frac{t^4}{\beta^2} \sum F_2(i\omega_l + i\omega_m - i\omega_n + i\omega_o - i\omega_p) F_3(i\omega_l + i\omega_m + i\omega_o) \times \\ F_2(i\omega_l + i\omega_o) c^\dagger_{k\alpha}(i\Omega) c^\dagger_{k'\beta}(i\omega_n) c_{k''\gamma}(i\omega_m) c_{k'''\delta}(i\omega_l) f^\dagger_{1,\tau}(i\omega_p) f_{1,\tau'}(i\omega_o) \,. \quad \text{(S-22)}$$

The sum runs over all frequencies and momenta. $i\Omega = i\omega_l + i\omega_m - i\omega_n + i\omega_o - i\omega_p$ ensures energy conservation. The sum over the orbital labels is set by the specific algebraic form of $\mathbf{A}_2$ and $\mathbf{w}$ in Eqs. (S-16) and (S-18) respectively. A mean-field treatment of the four point interaction $\left(c^\dagger c\right)^2$ close to the free propagator poles permits to map also this interaction onto Eq. (S-6). All correction to the mean-field and frequency expansion is irrelevant in the renormalization flow towards low energies and can be neglected. Lead electron operators in Eq. (S-22) are contracted by pairs and replaced by their free propagator. For instance

$$\overline{c^\dagger_{k\alpha}(i\Omega) c^\dagger_{k'\beta}(i\omega_n) c_{k''\gamma}}(i\omega_m) c_{k'''\delta}(i\omega_l) f^\dagger_{1,\tau}(i\omega_p) f_{1,\tau'}(i\omega_o) = \\ -G_k(i\omega_m) c^\dagger_{k'\beta}(i\omega_l + i\omega_o - i\omega_p) c_{k'''\delta}(i\omega_l) f^\dagger_{1,\tau}(i\omega_p) f_{1,\tau'}(i\omega_o) \,. \quad \text{(S-23)}$$

The propagators appearing in Eq. (S-22) are expanded close to the fixed point frequencies $i\omega \to E_F = 0$ for lead electrons and $i\tilde{\omega} \to \tilde{\varepsilon}_d \sim \varepsilon_d + \lambda$ for dot para-fermions, leading to contributions of the form

$$V = A_n \times \frac{t^4}{\beta} \sum c^\dagger_{k\nu}(i\omega_l + i\omega_o - i\omega_p) c_{k'\nu'}(i\omega_l) f^\dagger_{1,\tau}(i\omega_p) f_{1,\tau'}(i\omega_o) \,, \quad \text{(S-24)}$$

in which different contractions lead to

$$A_0 = \frac{F_2^2(i\omega + i\tilde{\omega})}{\beta} \sum_{p,n} G_p(i\omega_n) F_3(i\omega_n + i\tilde{\omega} + i\omega) = \frac{\nu_0}{(\varepsilon_d + U)^2} \ln\frac{2\varepsilon_d + 3U}{\Xi} \,,$$

$$A_1 = -\frac{F_2(i\omega + i\tilde{\omega})}{\beta} \sum_{p,i\omega_n} G_p(i\omega_n) F_3(i\omega_n + i\tilde{\omega} + i\omega) F_2(i\omega_n + i\tilde{\omega}) = \frac{\nu_0}{(\varepsilon_d + U)(\varepsilon_d + 2U)} \ln\frac{2\varepsilon_d + 3U}{\varepsilon_d + U} \,, \quad \text{(S-25)}$$

$$A_2 = \frac{1}{\beta} \sum_{p,i\omega_n} G_p(i\omega_n) F_3(i\omega_n + i\tilde{\omega} + i\omega) F_2^2(i\omega_n + i\tilde{\omega}) = \frac{-\nu_0}{(\varepsilon_d + U)(\varepsilon_d + 2U)} + \frac{\nu_0}{(\varepsilon_d + 2U)^2} \ln\frac{2\varepsilon_d + 3U}{\varepsilon_d + U} \,.$$

The Matsubara sums are carried out first and the analytical continuations $i\omega \to 0$ and $i\tilde{\omega} \to \varepsilon_d + \lambda$ are done later. The cutoff $\Xi$ must be introduced to prevent UV divergences and can be safely sent to infinity at the end of calculations. Applying the constraint $\sum_\tau f^\dagger_{1,\tau} f_{1,\tau} = 1$ and Eq. (S-7), Eq. (S-22) maps onto Eq. (S-6), namely

$$t^4 \sum_{kk'\nu\nu'\tau\tau'} \text{Tr}\left\{\left[(8A_1 - 4A_0 - 4A_2)\mathbf{T}_{\tau\tau'} \cdot \mathbf{t}_{\nu\nu'} + \left(\frac{3}{2}A_0 + \frac{3}{2}A_2 - 3A_1\right)\delta_{\nu\nu'}\delta_{\tau\tau'}\right] c^\dagger_{k\nu} c_{k'\nu'} f^\dagger_{1,\tau} f_{1,\tau'}\right\} \,. \quad \text{(S-26)}$$

The effective interaction of the SU(4) Anderson Hamiltonian recovers then Eq. (S-6) with couplings

$$\tilde{J}_{n,m,l} = \tilde{J}^1_{i\nu_n,i\omega_m,i\omega_l} + t^4 (8A_1 - 4A_0 - 4A_2) \,, \qquad \tilde{W}_{n,m,l} = \tilde{W}^1_{i\nu_n,i\omega_m,i\omega_l} + t^4 \left(\frac{3}{2}A_0 + \frac{3}{2}A_2 - 3A_1\right) \,. \quad \text{(S-27)}$$

### 2. Calculation of the slave-field propagator

To leading order in $t$, the self-energy $\Sigma_1(i\omega_n)$ is given by the diagram pictured in Fig. S-3

$$\Sigma_1(i\omega_n) = \frac{4}{\beta} \sum_{p,l} G_p(i\omega_l) \tilde{W}^1_{i\omega_l, i\omega_l, i\omega_n - i\omega_l} = \nu_0 t^2 \left[\ln\frac{\lambda - i\omega}{\Xi} + 3\ln\frac{2\varepsilon_d + U + \lambda - i\omega}{\Xi}\right] \,. \quad \text{(S-28)}$$

Notice that $\omega$ can be safely sent to zero at this stage. The anti-symmetric properties of the SU(N) matrices $\mathbf{T}$ and $\mathbf{t}$, imply that only $\tilde{W}^1$ in Eq. (S-21), and not $\tilde{J}^1$, contributes to the self energy. The full propagator reads then

$$\mathcal{F}_1(i\omega_n) = \frac{\mathcal{Z}}{i\omega_n - \tilde{\varepsilon}_d} \,, \quad \text{(S-29)}$$

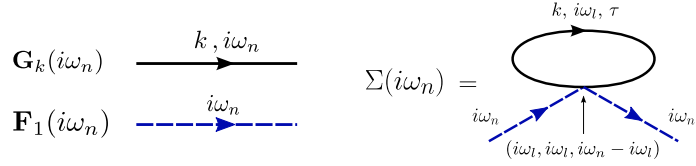

FIG. S-3: Left) Diagrammatic representation of free propagators. Solid lines are associated to lead electrons of momentum $k$ and frequency $i\omega_n$ and dashed lines to slave-fermions of charge 1. Right) One-loop contribution to the self-energy of the para-fermions. The arrows indicate the frequency dependence of the couplings in Eq. (S-21).

involving the renormalized pole $\tilde{\varepsilon}_d = \varepsilon_d + \lambda + \Sigma_1(\tilde{\varepsilon}_d)$ and the quasi-particle weight $\mathcal{Z} = 1/\left[1 - \Sigma'(\tilde{\varepsilon}_d)\right]$

$$\mathcal{Z} = 1 + \nu_0 t^2 \left[\frac{1}{\varepsilon_d} - \frac{3}{\varepsilon_d + U}\right], \qquad \tilde{\varepsilon}_d = \varepsilon_d + \lambda + \nu_0 t^2 \left(\ln\frac{-\varepsilon_d}{\Xi} + 3\ln\frac{\varepsilon_d + U}{\Xi}\right). \tag{S-30}$$

### 3. Calculation of the four-point vertex function

To fourth order in $t$, the calculation of the vertex Eq. (S-8) involves the diagrams in Fig. S-4. These are calculated using the $S_{\text{SU}(4)}$ interaction Eq. (S-6) with the couplings Eq. (S-27). The first contribution to the renormalized vertex $\mathcal{V}^R$ is given by the first diagram of the series in Fig. S-4 multiplied by the quasi-particle weight $\mathcal{Z}$ Eq. (S-30) and performing the analytical extension $i\tilde{\omega} \to \tilde{\varepsilon}_d$, given in Eq. (S-30). This contribution reads

$$\begin{aligned}
\mathcal{V}_1 &= \mathcal{Z}\left[\tilde{J}^1_{i\omega,i\omega,i\tilde{\omega}-i\omega}\mathbf{T}\cdot\mathbf{t} + \tilde{W}^1_{i\omega,i\omega,i\tilde{\omega}-i\omega}\mathbb{1}\right]\Big|_{i\omega\to 0,\; i\tilde{\omega}\to\tilde{\varepsilon}_d} \\
&= \mathbf{T}\cdot\mathbf{t}\left[J_1\mathcal{Z} + 2\nu_0 t^4\left(\ln\frac{-\varepsilon_d}{\Xi} + 3\ln\frac{\varepsilon_d+U}{\Xi}\right)\left(\frac{1}{\varepsilon_d^2} + \frac{1}{(\varepsilon_d+U)^2}\right)\right] + \\
&\quad \mathbb{1}\left[W_1\mathcal{Z} + \frac{\nu_0 t^4}{4}\left(\ln\frac{-\varepsilon_d}{\Xi} + 3\ln\frac{\varepsilon_d+U}{\Xi}\right)\left(\frac{1}{\varepsilon_d^2} - \frac{3}{(\varepsilon_d+U)^2}\right)\right],
\end{aligned} \tag{S-31}$$

with $J_1 = -2t^2 U/\varepsilon_d(\varepsilon_d + U)$ and $W_1 = -t^2(4\varepsilon_d + U)/4\varepsilon_d(\varepsilon_d + U)$, the couplings obtained by the Schrieffer-Wolff transformation. The leading diagram in Fig. S-4 also carries a subleading contribution from Eqs. (S-26) and (S-27). This reads

$$\mathcal{V}'_1 = t^2\left(8A_1 - 4A_0 - 4A_2\right)\mathbf{T}\cdot\mathbf{t} + t^2\left(\frac{3}{2}A_0 + \frac{3}{2}A_2 - 3A_1\right)\mathbb{1}. \tag{S-32}$$

The diagrams of order $t^4$ in Fig. S-4 are calculated and cast in the form

$$\begin{aligned}
\mathcal{V}^a &= -\frac{1}{\beta}\sum_{pi\omega_n\alpha\beta} G_p(i\omega + i\tilde{\omega} - i\omega_n)F_1(i\omega_n)\left[\left(-\frac{1}{4}J_a^2 + 2J_a W_a\right)\mathbf{T}\cdot\mathbf{t} + \left(\frac{15}{64}J_a^2 + W_a^2\right)\mathbb{1}\right], \\
\mathcal{V}^b &= -\frac{1}{\beta}\sum_{pi\omega_n\alpha\beta} G_p(i\omega + i\omega_n - i\tilde{\omega})F_1(i\omega_n)\left[\left(\frac{7}{4}J_b^2 + 2J_b W_b\right)\mathbf{T}\cdot\mathbf{t} + \left(\frac{15}{64}J_b^2 + W_b^2\right)\mathbb{1}\right],
\end{aligned} \tag{S-33}$$

with

$$\begin{aligned}
J_a &= -2t^2\left[F_0(i\omega_n - i\omega) + F_2(i\omega + i\tilde{\omega})\right], & W_a &= -\frac{t^4}{4}\left[F_0(i\omega_n - i\omega) - 3F_2(i\omega + i\tilde{\omega})\right], \\
J_b &= -2t^2\left[F_0(i\tilde{\omega} - i\omega) + F_2(i\omega + i\omega_n)\right], & W_b &= -\frac{t^2}{4}\left[F_0(i\tilde{\omega} - i\omega) - 3F_2(i\omega + i\omega_n)\right].
\end{aligned} \tag{S-34}$$

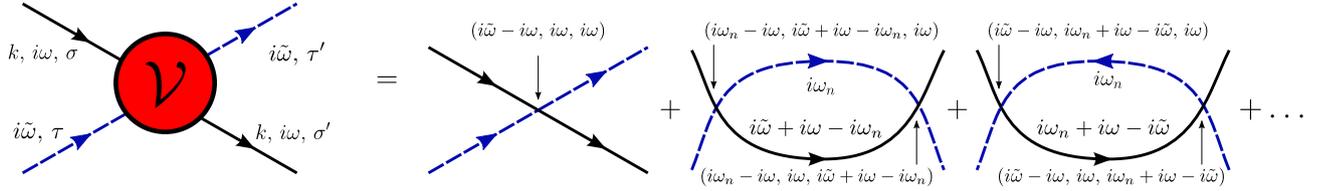

FIG. S-4: Diagrammatic series for the vertex function Eq. (S-8). This is a more detailed version of Fig. 3 in the man text. The arrows indicate the frequency dependence of the couplings in Eq. (S-21).

The following integrals are computed

$$\begin{aligned}
I_{00} &= \frac{1}{\beta}\sum_{p,n} G_p(i\omega+i\tilde\omega-i\omega_n)F_1(i\omega_n)F_0^2(i\omega_n-i\omega) = \frac{\nu_0}{\varepsilon_d^2}\left[\ln\frac{-\varepsilon_d}{\omega}-1\right], \\
I_{02} &= \frac{1}{\beta}\sum_{p,n} G_p(i\omega+i\tilde\omega-i\omega_n)F_1(i\omega_n)F_0(i\omega_n-i\omega)F_2(i\omega+i\tilde\omega) = \frac{\nu_0}{\varepsilon_d(\varepsilon_d+U)}\ln\frac{\omega}{-\varepsilon_d}, \\
I_{22} &= \frac{1}{\beta}\sum_{p,n} G_p(i\omega+i\tilde\omega-i\omega_n)F_1(i\omega_n)F_2^2(i\omega+i\tilde\omega) = -\frac{\nu_0}{(\varepsilon_d+U)^2}\ln\frac{\omega}{\Xi}, \\
L_{00} &= \frac{1}{\beta}\sum_{p,n} G_p(i\omega+i\omega_n-i\tilde\omega)F_1(i\omega_n)F_0^2(i\tilde\omega-i\omega) = \frac{\nu_0}{\varepsilon_d^2}\ln\frac{\omega}{\Xi}, \\
L_{02} &= \frac{1}{\beta}\sum_{p,n} G_p(i\omega+i\omega_n-i\tilde\omega)F_1(i\omega_n)F_0(i\tilde\omega-i\omega)F_2(i\omega+i\omega_n) = \frac{\nu_0}{\varepsilon_d(\varepsilon_d+U)}\ln\frac{\varepsilon_d+U}{\omega}, \\
L_{22} &= \frac{1}{\beta}\sum_{p,n} G_p(i\omega+i\omega_n-i\tilde\omega)F_1(i\omega_n)F_2^2(i\omega+i\omega_n) = \frac{\nu_0}{(\varepsilon_d+U)^2}\left[1-\ln\frac{\varepsilon_d+U}{\omega}\right].
\end{aligned} \quad (\text{S-35})$$

In this case, one has to take care of the infra-red cutoff $\omega$ for sums over momenta. Only the integrals $I_{22}$ and $L_{00}$ depend on the high energy cut-off $\Xi$. The contributions to the vertex function read then

$$\begin{aligned}
\mathcal{V}^a &= t^4(4I_{02}+4I_{22})\mathbf{T}\cdot\mathbf{t}-t^4\left(I_{00}+\frac{3}{2}I_{22}+\frac{3}{2}I_{02}\right)\mathbb{1}, \\
\mathcal{V}^b &= -t^4(8L_{00}+12L_{02}+4L_{22})\mathbf{T}\cdot\mathbf{t}-t^4\left(L_{00}+\frac{3}{2}L_{22}+\frac{3}{2}L_{02}\right)\mathbb{1}.
\end{aligned} \quad (\text{S-36})$$

The summation of Eq. (S-31), (S-32) and (S-36) leads to the final expression of the renormalized vertex in the form Eq. (S-10). The $\Xi\to\infty$ limit can be safely performed. The spin-exchange part $\mathcal{J}(\omega)$ recovers Eq. (7a) in the main text. For completeness, we also report the result for the potential scattering term

$$\mathcal{W}_1 = W_1 + \frac{1}{\nu_0}\left(\frac{\Delta}{\pi U}\right)^2 \eta_1\left(\frac{\varepsilon_d}{U}\right), \quad (\text{S-37})$$

with

$$\begin{aligned}
\eta_1(x) =& \frac{3}{4}\left(\frac{1}{x^2}+\frac{1}{(x+1)^2}\right)+\frac{1}{x^2}\ln\frac{\sqrt[4]{-x(x+1)}}{-x}+\frac{3}{(x+1)^2}\ln\frac{\sqrt{(x+1)(2x+3)}}{\sqrt[4]{-x(x+1)}}+ \\
& \frac{3}{2}\frac{1}{x(x+1)}\ln\frac{-x}{x+1}+3\left[\frac{1}{(x+2)^2}\ln\sqrt{\frac{2x+3}{x+1}}-\frac{1}{(x+1)(x+2)}\left(\frac{1}{2}+\ln\frac{2x+3}{x+1}\right)\right].
\end{aligned} \quad (\text{S-38})$$

Notice that even if $W_1 = -t^2(4\varepsilon_d+U)/4\varepsilon_d(\varepsilon_d+U)$ vanishes in the middle of the Coulomb valley for $\varepsilon_d = -U/2$, $\eta_1$ does not. This is explained by the fact that the model is effectively particle-hole symmetric at $\varepsilon_d = -U/2$ only to second order in $t$.

# S-III. CALCULATION OF THE SU(4) SPIN EXCHANGE PROCESSES FOR $q = 2$

### 4. Integration of the high energy slave-fields

For $-2U < \varepsilon_d < -U$, two charges are frozen on the dot to low energy. To fourth order in the tunneling $t$, all the other occupancies of the quantum dot contribute virtually to the effective action, see Fig. S-2. The highest energy charge modes are $b_0$ and $b_4$. They are integrated first and the effective action reads

$$S_2 = -\text{Tr}\left[\mathbf{b}^\dagger \cdot F_2^{-1}\mathbb{1}_6 \cdot \mathbf{b}\right] - \text{Tr}\left[\mathbf{f}^\dagger \cdot \Phi_1^{-1} \cdot \mathbf{f}\right] - \text{Tr}\left[\mathbf{F}^\dagger \cdot \Phi_3^{-1} \cdot \mathbf{F}\right] + t\,\text{Tr}\left[\mathbf{D}_1^\dagger \cdot \mathbf{f} + \mathbf{f}^\dagger \cdot \mathbf{B}_1 + \mathbf{B}_3^\dagger \cdot \mathbf{F} + \mathbf{F}^\dagger \cdot \mathbf{B}_3\right]. \quad \text{(S-39)}$$

The vectors $\mathbf{F}$ and $\mathbf{b}$ are the same as in Eq. (S-15) and we introduce

$$\mathbf{B}_1 = \begin{pmatrix} c_2^\dagger b_{12} + c_3^\dagger b_{13} + c_4^\dagger b_{14} \\ -c_1^\dagger b_{12} + c_3^\dagger b_{23} + c_4^\dagger b_{24} \\ -c_1^\dagger b_{13} - c_2^\dagger b_{23} + c_4^\dagger b_{34} \\ -c_1^\dagger b_{14} - c_2^\dagger b_{24} - c_3^\dagger b_{34} \end{pmatrix}, \quad \mathbf{B}_3 = \begin{pmatrix} b_{34}c_2 - b_{24}c_3 + b_{23}c_4 \\ -b_{34}c_1 + b_{14}c_3 - b_{13}c_4 \\ b_{24}c_1 - b_{14}c_2 + b_{12}c_4 \\ -b_{23}c_1 + b_{13}c_2 - b_{12}c_3 \end{pmatrix}, \quad \mathbf{f} = \begin{pmatrix} f_{1,1} \\ f_{1,2} \\ f_{1,3} \\ f_{1,4} \end{pmatrix},$$

$$\Phi_1^{-1} = F_1^{-1}\mathbb{1}_4 - t^2\mathbf{A}_1, \qquad \Phi_3^{-1} = F_3^{-1}\mathbb{1}_4 - t^2\mathbf{A}_3, \quad \text{(S-40)}$$

$$\mathbf{A}_1 = \begin{pmatrix} c_1 \\ c_2 \\ c_3 \\ c_4 \end{pmatrix} \cdot F_0\mathbb{1}_4 \cdot \begin{pmatrix} c_1^\dagger & c_2^\dagger & c_3^\dagger & c_4^\dagger \end{pmatrix}, \quad \mathbf{A}_3 = \begin{pmatrix} c_1^\dagger \\ c_2^\dagger \\ c_3^\dagger \\ c_4^\dagger \end{pmatrix} \cdot F_4\mathbb{1}_4 \cdot \begin{pmatrix} c_1 & c_2 & c_3 & c_4 \end{pmatrix}.$$

The integration of the remaining para-fermions collected in the vectors $\mathbf{f}$ and $\mathbf{F}$ is also Gaussian

$$S_2' = -\text{Tr}\left[\mathbf{b}^\dagger \cdot F_2^{-1} \cdot \mathbf{b}\right] + \text{Tr}\left[\tilde{J}^2\left(\mathbf{b}^\dagger \cdot \mathbf{T} \cdot \mathbf{b}\right)\cdot\left(\mathbf{c}^\dagger \cdot \mathbf{t} \cdot \mathbf{c}\right) + \tilde{W}^2\left(\mathbf{b}^\dagger \cdot \mathbb{1}_6 \cdot \mathbf{b}\right)\left(\mathbf{c}^\dagger\mathbb{1}_4\mathbf{c}\right)\right]$$
$$+ t^4\text{Tr}\left[\mathbf{B}_1^\dagger \cdot F_1 \cdot \mathbf{A}_1 \cdot F_1 \cdot \mathbf{B}_1\right] + t^4\text{Tr}\left[\mathbf{B}_3^\dagger \cdot F_3 \cdot \mathbf{A}_3 \cdot F_3 \cdot \mathbf{B}_3\right], \quad \text{(S-41)}$$

with this time $\tilde{J}^2 = -2t^2[F_1 + F_3]$ and $\tilde{W}^2 = t^2/2[F_3 - F_1]$. In the Matsubara representation

$$\tilde{J}^2_{i\nu_n, i\omega_m, i\omega_l} = -2t^2\left[F_1(i\nu_n - i\omega_l) + F_3(i\nu_n + i\omega_m)\right], \quad \tilde{W}^2_{i\nu_n, i\omega_m, i\omega_l} = -\frac{t^2}{2}\left[F_1(i\nu_n - i\omega_l) - F_3(i\nu_n + i\omega_m)\right]. \quad \text{(S-42)}$$

Taking the analytical continuations $i\nu \to \varepsilon_{d2} = 2\varepsilon_d + U + \lambda$ and $i\omega_{1,2} \to 0$, also these expressions recover the Schrieffer-Wolff couplings $J_2 = -2t^2U/(\varepsilon_d + U)(\varepsilon_d + 2U)$ and $W_2 = -t^2(2\varepsilon_d + 3U)/4(\varepsilon_d + U)(\varepsilon_d + 2U)$, valid for $q = 2$. The SU(4) representation of the interaction $(\mathbf{b}^\dagger \cdot \mathbf{T} \cdot \mathbf{b})(\mathbf{c}^\dagger \cdot \mathbf{t} \cdot \mathbf{c})$ deserves a specific discussion. The vector $\mathbf{b}$ has six components. This implies that the matrices composing the invariant SU(4) vector $\mathbf{T}$ are not those of the fundamental representation [3] and present in Eq. (S-7). The explicit expression of the product in Eq. (S-41) reads

$$2\left(\mathbf{b}^\dagger \cdot \mathbf{T} \cdot \mathbf{b}\right)\left(\mathbf{c}^\dagger \cdot \mathbf{t} \cdot \mathbf{c}\right) = \left[c_1^\dagger c_2\left(b_{24}^\dagger b_{14} + b_{23}^\dagger b_{13}\right) + c_1^\dagger c_3\left(b_{34}^\dagger b_{14} - b_{23}^\dagger b_{12}\right) + c_1^\dagger c_4\left(-b_{34}^\dagger b_{13} - b_{24}^\dagger b_{12}\right) + \right.$$
$$\left. c_2^\dagger c_3\left(b_{34}^\dagger b_{24} + b_{13}^\dagger b_{12}\right) + c_2^\dagger c_4\left(-b_{34}^\dagger b_{23} + b_{14}^\dagger b_{12}\right) + c_3^\dagger c_4\left(b_{24}^\dagger b_{23} + b_{14}^\dagger b_{13}\right) + \text{h.c.}\right] +$$
$$\frac{1}{2}c_1^\dagger c_1\left(b_{12}^\dagger b_{12} + b_{13}^\dagger b_{13} + b_{14}^\dagger b_{14} - b_{23}^\dagger b_{23} - b_{24}^\dagger b_{24} - b_{34}^\dagger b_{34}\right) +$$
$$\frac{1}{2}c_2^\dagger c_2\left(b_{12}^\dagger b_{12} + b_{23}^\dagger b_{23} + b_{24}^\dagger b_{24} - b_{13}^\dagger b_{13} - b_{14}^\dagger b_{14} - b_{34}^\dagger b_{34}\right) +$$
$$\frac{1}{2}c_3^\dagger c_3\left(b_{13}^\dagger b_{13} + b_{23}^\dagger b_{23} + b_{34}^\dagger b_{34} - b_{12}^\dagger b_{12} - b_{14}^\dagger b_{14} - b_{24}^\dagger b_{24}\right) +$$
$$\frac{1}{2}c_4^\dagger c_4\left(b_{14}^\dagger b_{14} + b_{24}^\dagger b_{24} + b_{34}^\dagger b_{34} - b_{12}^\dagger b_{12} - b_{13}^\dagger b_{13} - b_{23}^\dagger b_{23}\right). \quad \text{(S-43)}$$

This is derived by applying the constraint $\sum_{\tau' > \tau} b^\dagger_{\tau\tau'} b_{\tau\tau'} = 1$, which is the corrected version of Eq. (S-3), once all the other slave-fields have been integrated. The mapping onto the fermionic representation of the SU(4) spin-exchange interaction Eq. (S-6) is obtained refermionizing the slave-bosons $b_{\tau\tau'}$. These fields are associated to a two-electron(fermion) state $b^\dagger_{\tau\tau'} = f_\tau^\dagger f_{\tau'}^\dagger$. The $f_\tau$ are para-fermion fields obeying the constraint $\sum_\tau f_\tau^\dagger f_\tau = 2$. Substituting in Eq. (S-43), Eq. (S-7) for $q = 2$ is recovered, according to the effective SU(4) symmetric interaction Eq. (S-6).

The subleading interaction in Eq. (S-41) is simplified with a mean-field treatment, analog to that carried out in Section S-II 1, leading to

$$S_{t^4} = t^4 \sum_{kk'\nu\nu'\tau\tau'\rho\rho'} \text{Tr}\left\{\left[2(\mathcal{V}_0 - \mathcal{V}_4)\mathbf{T}_{\tau\tau',\rho\rho'} \cdot \mathbf{t}_{\nu\nu'} + \frac{\mathcal{V}_0 + \mathcal{V}_4}{2}\delta_{\tau\tau',\rho\rho'}\delta_{\nu\nu'}\right]c_{k\nu}^\dagger c_{k'\nu'}b_{\tau\tau'}^\dagger b_{\rho\rho'}\right\}. \quad \text{(S-44)}$$

$\mathcal{V}_0$ and $\mathcal{V}_4$ are the contributions to the vertex coming from the 0 and 4 charge sector respectively. They read

$$\begin{aligned}\mathcal{V}_0 &= B_0 + B_2 - 2B_1, \\ \mathcal{V}_4 &= C_0 + C_2 - 2C_1,\end{aligned} \quad \text{(S-45)}$$

with

$$\begin{aligned}
B_0 &= \frac{1}{\beta}\sum_{p,n} G_p(-i\omega_n)F_1^2(i\nu - i\omega)F_0(i\nu - i\omega + i\omega_n) = -\frac{\nu_0}{(\varepsilon_d + U)^2}\ln\frac{-2\varepsilon_d - U}{\Xi}, \\
B_1 &= \frac{-1}{\beta}\sum_{p,n} G_p(-i\omega_n)F_1(i\nu + i\omega_n)F_0(i\nu - i\omega + i\omega_n)F_1(i\nu - i\omega) = \frac{\nu_0}{\varepsilon_d(\varepsilon_d + U)}\ln\frac{\varepsilon_d + U}{2\varepsilon_d + U}, \\
B_2 &= \frac{1}{\beta}\sum_{p,n} G_p(-i\omega_n)F_1^2(i\nu + i\omega_n)F_0(i\nu - i\omega + i\omega_n) = \frac{\nu_0}{\varepsilon_d(\varepsilon_d + U)} + \frac{\nu_0}{\varepsilon_d^2}\ln\frac{\varepsilon_d + U}{2\varepsilon_d + U}, \\
C_0 &= \frac{1}{\beta}\sum_{p,n} G_p(i\omega_n)F_3^2(i\nu + i\omega)F_0(i\nu + i\omega + i\omega_n) = \frac{\nu_0}{(\varepsilon_d + 2U)^2}\ln\frac{2\varepsilon_d + 5U}{\Xi}, \\
C_1 &= -\frac{1}{\beta}\sum_{p,n} G_p(i\omega_n)F_1(i\nu + i\omega_n)F_0(i\nu + i\omega + i\omega_n)F_1(i\nu + i\omega) = \frac{\nu_0}{(\varepsilon_d + 3U)(\varepsilon_d + 2U)}\ln\frac{2\varepsilon_d + 5U}{\varepsilon_d + 2U}, \\
C_2 &= \frac{1}{\beta}\sum_{p,n} G_p(i\omega_n)F_3^2(i\nu + i\omega_n)F_0(i\nu + i\omega + i\omega_n) = -\frac{\nu_0}{(\varepsilon_d + 3U)(\varepsilon_d + 2U)} + \frac{\nu_0}{(\varepsilon_d + 3U)^2}\ln\frac{2\varepsilon_d + 5U}{\varepsilon_d + 2U}.
\end{aligned} \quad \text{(S-46)}$$

The total effective interaction maps onto the Kondo-like interaction Eq. (S-6) with effective couplings

$$\tilde{J}_{n,m,l} = \tilde{J}^2_{i\nu_n,i\omega_m,i\omega_l} + t^2(\mathcal{V}_0 - \mathcal{V}_4), \qquad \tilde{W}_{n,m,l} = \tilde{W}^2_{i\nu_n,i\omega_m,i\omega_l} + \frac{t^4}{2}(\mathcal{V}_0 + \mathcal{V}_4). \quad \text{(S-47)}$$

### 5. Calculation of the slave-field propagator

To second order in $t$, the self-energy of the boson propagator $\mathcal{F}_2(i\nu_n) = -\langle b(i\nu_n)b^\dagger(i\nu_n)\rangle$ reads

$$\Sigma_2(i\nu_n) = \frac{4}{\beta}\sum_{p,l} W(i\nu_n, i\omega_l, i\omega_l)G_p(i\omega_l) = \frac{2\Delta}{\pi}\left[\ln\frac{-i\nu_n + 3\varepsilon_d + 3U + \lambda}{\Xi} + \ln\frac{\varepsilon_d + \lambda - i\nu_n}{\Xi}\right]. \quad \text{(S-48)}$$

The self-energy leads to the renormalized pole $i\tilde{\nu}$ and quasi-particle weight $\mathcal{Z}_2$

$$i\tilde{\nu} \to \tilde{\varepsilon}_{d2} = 2\varepsilon_d + U + \lambda + 2\frac{\Delta}{\pi}\left[\ln\frac{\varepsilon_d + 2U}{\Xi} + \ln\frac{-\varepsilon_d - U}{\Xi}\right], \qquad \mathcal{Z}_2 = 1 - \nu_0 J_2, \quad \text{(S-49)}$$

with $J_2 = -2t^2 U/(\varepsilon_d + U)(\varepsilon_d + 2U)$ given by the Schrieffer-Wolff transformation.

### 6. Calculation of the four-point vertex function

Although the possibility of the fermionic representation of the slave-boson fields, it is much simpler to calculate the vertex function for the slave-bosons, similarly to Eq. (S-8)

$$\mathcal{V}^\omega_{\nu\nu';\tau\tau',\rho\rho'}(k, i\omega_l, i\nu_m; k'i\omega_n, i\nu_o) = -\beta\frac{\langle c_{k\nu}(i\omega_l)b_{\tau\tau'}(i\nu_m)b^\dagger_{\rho\rho'}(i\nu_o)c^\dagger_{k'\nu'}(i\omega_n)\rangle\big|_c}{G_k(i\omega_l)G_{k'}(i\omega_n)\mathcal{F}_2(i\nu_n)\mathcal{F}_2(i\nu_o)}. \quad \text{(S-50)}$$

This takes the form Eq. (S-10), but the vector **T** is made of the 6 × 6 matrices appearing in Eq. (S-41), leading to Eq. (S-43). We calculate now all the contributions to the renormalized vertex $\mathcal{V}_2^R = \mathcal{Z}_2 \mathcal{V}^\omega$. The contribution corresponding to the first diagram of the series in Fig. S-4 is readily obtained

$$\mathcal{V}_{1-2} = \mathbf{T} \cdot \mathbf{t} \left[ J_2 - \nu_0 J_2^2 + 2\nu_0 \left( \frac{J_2^2}{4} + 4W_2^2 \right) \ln \frac{(-\varepsilon_d - U)(\varepsilon_d + 2U)}{\Xi^2} \right] + \tag{S-51}$$
$$\mathbb{1} \left[ W_2 - \nu_0 J_2 W_2 + \nu_0 J_2 W_2 \ln \frac{(-\varepsilon_d - U)(\varepsilon_d + 2U)}{\Xi^2} \right].$$

This includes the corrections to the pole and the quasi-particle weight Eq. (S-49). Also the contributions coming from the subleading interaction Eq. (S-44) have to be taken into account

$$\mathcal{V}_2' = 2t^4 \left( \mathcal{V}_0 - \mathcal{V}_4 \right) \mathbf{T} \cdot \mathbf{t} + \frac{t^4}{2} \left( \mathcal{V}_0 + \mathcal{V}_2 \right) \mathbb{1}. \tag{S-52}$$

Eventually, we have the contributions which are the analog of the contributions $\mathcal{V}^a$ and $\mathcal{V}^b$ in Eq. (S-33)

$$\mathcal{V}_2^a = -\frac{1}{\beta} \sum_{p,n} G_p(i\omega + i\nu_\omega - i\nu_n) F_2(i\nu_n) \left[ \left( -J_a^2 + 2J_a W_a \right) \mathbf{T} \cdot \mathbf{t} + \left( \frac{5}{16} J_a^2 + W_a^2 \right) \mathbb{1} \right],$$
$$\mathcal{V}_2^b = -\frac{1}{\beta} \sum_{p,n} G_p(i\omega + i\nu_n - i\nu_\omega) F_2(i\nu_n) \left[ \left( J_b^2 + 2J_b W_b \right) \mathbf{T} \cdot \mathbf{t} + \left( \frac{5}{16} J_b^2 + W_b^2 \right) \mathbb{1} \right], \tag{S-53}$$

with

$$J_a = -2t^2 \left[ F_1(i\nu_n - i\omega) + F_3(i\omega + i\nu_\omega) \right], \quad W_a = -\frac{t^4}{2} \left[ F_1(i\nu_n - i\omega) - F_3(i\omega + i\nu_\omega) \right],$$
$$J_b = -2t^2 \left[ F_1(i\nu_\omega - i\omega) + F_3(i\omega + i\nu_n) \right], \quad W_b = -\frac{t^2}{2} \left[ F_1(i\nu_\omega - i\omega) - F_3(i\omega + i\nu_n) \right]. \tag{S-54}$$

The calculation of the Matsubara sums and of the integrals over momenta leads to the following result

$$\mathcal{V}_2^a = 2t^4 \left( I_{11} + 3I_{33} + 4I_{13} \right) \mathbf{T} \cdot \mathbf{t} - \frac{t^4}{2} \left( 3I_{11} + 3I_{33} + 4I_{13} \right) \mathbb{1},$$
$$\mathcal{V}_2^b = -2t^4 \left( 3L_{11} + L_{33} + 4L_{13} \right) \mathbf{T} \cdot \mathbf{t} - \frac{t^4}{2} \left( 3L_{11} + 3L_{33} + 4L_{13} \right) \mathbb{1}, \tag{S-55}$$

with

$$I_{11} = \frac{1}{\beta} \sum_{p,n} F_2(i\nu_n) G_p(i\omega + i\nu_\omega - i\nu_n) F_1^2(i\nu_n - i\omega) = -\frac{\nu_0}{(\varepsilon_d + U)^2} \left[ 1 + \ln \frac{\omega}{-\varepsilon_d - U} \right],$$
$$I_{13} = \frac{1}{\beta} \sum_{p,n} F_2(i\nu_n) G_p(i\omega + i\nu_\omega - i\nu_n) F_1(i\nu_n - i\omega) F_3(i\omega + i\nu_\omega) = \frac{\nu_0}{(\varepsilon_d + U)(\varepsilon_d + 2U)} \ln \frac{\omega}{-\varepsilon_d - U},$$
$$I_{33} = \frac{1}{\beta} \sum_{p,n} F_2(i\nu_n) G_p(i\omega + i\nu_\omega - i\nu_n) F_3^2(i\omega + i\nu_\omega) = -\frac{\nu_0}{(\varepsilon_d + 2U)^2} \ln \frac{\omega}{\Xi},$$
$$L_{11} = \frac{1}{\beta} \sum_{p,n} F_2(i\nu_n) G_p(i\omega + i\nu_n - i\nu_\omega) F_1^2(i\nu_\omega - i\omega) = \frac{\nu_0}{(\varepsilon_d + U)^2} \ln \frac{\omega}{\Xi},$$
$$L_{13} = \frac{1}{\beta} \sum_{p,n} F_2(i\nu_n) G_p(i\omega + i\nu_n - i\nu_\omega) F_1(i\nu_\omega - i\omega) F_3(i\omega + i\nu_n) = -\frac{\nu_0}{(\varepsilon_d + U)(\varepsilon_d + 2U)} \ln \frac{\omega}{\varepsilon_d + 2U},$$
$$L_{33} = \frac{1}{\beta} \sum_{p,n} F_2(i\nu_n) G_p(i\omega + i\nu_n - i\nu_\omega) F_3^2(i\omega + i\nu_n) = \frac{\nu_0}{(\varepsilon_d + 2U)^2} \left[ 1 + \ln \frac{\omega}{\varepsilon_d + 2U} \right]. \tag{S-56}$$

The total expression of the renormalized vertex is obtained by summing up Eqs. (S-51), (S-55) and (S-52). The limit $\Xi \to \infty$ is safely carried out. The renormalized vertex recovers Eq. (S-10) and the spin-exchange coupling reads

$$\mathcal{J}_2(\omega) = J_2 + 2\nu_0 J_2^2 g_2 \left( \frac{\varepsilon_d}{U} \right) - 2\nu_0 J_2^2 \ln \frac{\omega}{e^{-\frac{1}{2}} \sqrt{(-\varepsilon_d - U)(\varepsilon_d + 2U)}}, \tag{S-57}$$

with

$$g_2(x) = \frac{(x+2)^2}{4x^2}\left[x + \ln\frac{x+1}{2x+1}\right] + \frac{(x+1)^2}{4(x+3)^2}\left[-x - 3 + \ln\frac{x+2}{2x+5}\right]. \tag{S-58}$$

Eq. (S-57) maps onto Eq.(2) in the main text if $J \to J_2 + 2\nu_0 J_2^2 g_2\left(\frac{\varepsilon_d}{U}\right)$ and $\mathcal{D} \to \mathcal{D}_2 = e^{-\frac{1}{2}}\sqrt{(-\varepsilon_d - U)(\varepsilon_d + 2U)}$. The SU(4) Kondo temperature for $q = 2$ is derived in the form Eq. (3), with the prefactor $f_2(x)$

$$f_2(x) = \sqrt[4]{\frac{(-1-x)(x+2)}{e^2}} e^{g_2(x)}. \tag{S-59}$$

This function is plotted in the main text in Fig. 2, showing an excellent agreement with the numerical renormalization group calculations. For completeness we also give the potential scattering term

$$\mathcal{W}_2 = W_2 + \frac{1}{\nu_0}\left(\frac{\Delta}{\pi U}\right)^2 \eta_2\left(\frac{\varepsilon_d}{U}\right), \tag{S-60}$$

with

$$\begin{aligned}\eta_2(x) =& \frac{1}{2}\frac{2x+1}{x(x+1)^2} - \frac{1}{2}\frac{2x+5}{(x+3)(x+2)^2} - \frac{2}{(x+1)(x+2)}\ln\frac{x+2}{-x-1} + \frac{1}{2}\frac{1-x}{(x+1)x^2}\ln\frac{x+1}{2x+1} + \\ & \frac{1}{2}\frac{x+4}{(x+2)(x+3)^2}\ln\frac{x+2}{2x+5} + \frac{1}{(x+1)^2}\ln\frac{x+2}{\sqrt{(-x-1)(-2x-1)}} - \frac{1}{(x+2)^2}\ln\frac{-x-1}{\sqrt{(x+2)(2x+5)}}.\end{aligned} \tag{S-61}$$

$\mathcal{W}_2$ Eq. (S-60) vanishes at the particle-hole symmetric point $x = \varepsilon_d/U = -3/2$. This coincides with the middle of the Coulomb valley, in contrast with the $q = 1$ case, see Eq. (S-37).

## S-IV. CALCULATION OF THE SU(4) SPIN EXCHANGE PROCESSES FOR $q = 3$

The case of triple dot occupancy requires $-3U < \varepsilon_d < -2U$. The $q = 3$ sector is the particle-hole symmetric to $q = 1$. All results are readily derived by making the following set of substitutions in Section S-II

$$\begin{aligned}E_0 - E_1 &= -\varepsilon_d & \to & & E_4 - E_3 &= \varepsilon_d + 2U, \\ E_2 - E_1 &= \varepsilon_d + U & \to & & E_2 - E_3 &= -\varepsilon_d - 3U, \\ E_3 - E_1 &= 2\varepsilon_d + 3U & \to & & E_1 - E_3 &= -2\varepsilon - 3U.\end{aligned} \tag{S-62}$$

The prefactor for the SU(4) Kondo temperature, Eq. (3) in the main text, reads then $f_3(\varepsilon_d/U) = f_1(-2 - \varepsilon_d/U)$. The derivation of the effective SU(4) symmetric action Eq. (S-6) deserves a specific discussion. We recall that the constraint Eq. (S-3), imposing $N_{\text{slave}} = 1$, must be adapted upon the integration of the high energy slave-fields. In this case $\sum_\tau f_{3,\tau}^\dagger f_{3,\tau} = 1$. In this representation, the spin-exchange interaction reads

$$2\mathbf{T} \cdot \mathbf{t} = -\sum_{\tau\tau'} c_{k\tau}^\dagger c_{k'\tau'}\left(f_{3,\tau}^\dagger f_{3,\tau'} - \frac{1}{4}\delta_{\tau\tau'}\right). \tag{S-63}$$

This expression recovers the canonical one Eq. (S-7) with $q = 3$, if we substitute $f_{3,\tau}^\dagger f_{3,\tau'} = -f_{\tau'}^\dagger f_\tau$, in which these last fermionic fields obey the constraint $\sum_\tau f_\tau^\dagger f_\tau = 3$.

## S-V. NUMERICAL RENORMALIZATION GROUP APPROACH

In this section we provide some details regarding the numerical renormalization group (NRG) calculations. Our NRG computations were performed with a discretization parameter $\Lambda = 2$, while exploiting the full SU(4) symmetry of the Hamiltonian (with 300 kept multiplets corresponding to $\sim 4200$ kept states in each iteration). These calculations were checked against computations performed by using the four U(1)×U(1)×SU(2)×SU(2) symmetries with 3000 multiplets kept. [15]

The value of the Kondo temperature [16] depends on its precise definition and the quantity from which it is extracted. To determine it, we computed the spectral function of the $d$-level, $\mathcal{A}(\omega) = -\text{Im}\, G_{d_\tau d_\tau^\dagger}(\omega)/\pi$, and defined $T_{K,SU(4)}^{NRG}$ as the half-width at half maximum of $\mathcal{A}(\omega)$. This definition takes into account that the peak of the spectral function is generally not centered at the Fermi energy (excepting the electron-hole symmetrical point, $\epsilon_d = -3U/2$) and shifts gradually as a function of $\epsilon_d$. This definition is found to agree perfectly with the analytical expression, Eq. (5), up to a prefactor of the order one, $T_{K,SU(4)}^{NRG} \approx 0.94\, T_{K,SU(4)}$. In the NRG calculations, we changed $\Delta$ and $\epsilon_d$ simultaneously such that the size of the coupling $\nu_0 J_0$ remained constant, and extracted the $\epsilon_d$ dependent prefactor this way.

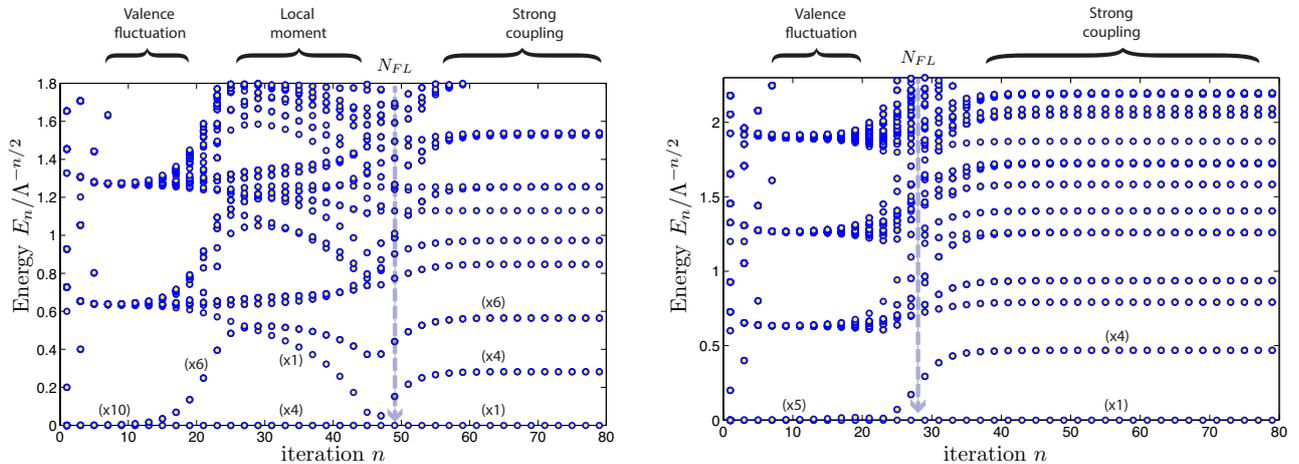

FIG. S-5: NRG finite size spectrum at T=0, and for the odd iterations. The iteration number $N_{FL}$ is indicated by an arrow and the degeneracies of the lowest lying states are indicated between the brackets. The parameters are: $\Lambda = 2$, $U = 0.15D$, $\Delta/U = 3 \times 10^{-4}$, $\varepsilon_d/U = -0.999$ (left panel) and $\varepsilon_d/U = -10^{-3}$ (right panel).

The various fixed points discussed in the main text can be identified from the flow diagram of the NRG levels [17, 18], shown in Fig. S-5 for parameters close to the mixed valence regime, $\varepsilon_d \approx \{0, -U\}$. For $\varepsilon_d = -U$, the ground state of the isolated dot ($\Delta = 0$) would be 10 fold degenerate. For finite but small enough $\Delta$, valence fluctuations mix these 10 states by quantum fluctuations and a valence fluctuation regime emerges. A transition to this valence fluctuation regime is expected therefore for $\epsilon_d \approx -U$ as soon as the energy drops below $U$. The interaction $U$ being relatively large, this mixed valence regime and the corresponding 10-fold ground state degeneracy is almost immediately observed in the finite size spectrum, after only a few iterations. Apart from the curious ground state degeneracy, the observed mixed valence spectrum displays a regular Fermi liquid-like structure, characteristic of the SU(4)-symmetrical mixed valence fixed point.

For small $\Delta$'s, (and $|\epsilon_d|, |\epsilon_d+U| \gg \Delta$) a second transition can be observed in the spectrum, occurring approximately at the charging energy, $\Delta E_q \approx \min_q |\epsilon_q|$, slightly shifted by mixed valence fluctuations. The ground state is gradually split and its degeneracy is partially removed when the energy drops below $\Delta E_q$, and the local moment regime develops. For $\varepsilon_d/U = -0.999$ and $\Delta/U = 3 \times 10^{-4}$, e.g., a fourfold degenerate $q = 1$ state prevails, and a corresponding 4-fold degenerate ground state emerges in the finite size spectrum for intermediate iteration numbers. Decreasing further the energy scale (increasing the iteration number), a Fermi liquid state develops below the Kondo scale, $T_{FL}$, where the ground state degeneracy is removed, and a phase shifted SU(4) Fermi liquid spectrum appears.

Getting closer to the valence transition point (i.e., for $|\epsilon_d|, |\epsilon_d+U| < \Delta$), the Kondo temperature gradually increases, the local moment regime shrinks, and at some point it disappears. This happens on the right panel of Fig. S-5, e.g., where one enters the Fermi liquid state directly from the mixed valence region.